\documentclass[10pt,conference]{IEEEtran}

\usepackage{algpseudocode}
\usepackage{tikz}
\usetikzlibrary{arrows.meta,positioning,calc,backgrounds}
\usepackage{caption}
\usepackage[dvipsnames]{xcolor}
\usepackage{booktabs}
\usepackage{tabularx}
\usepackage{colortbl}
\usepackage{arydshln}
\usepackage{adjustbox}
\usepackage{placeins}
\usepackage[ruled,vlined]{algorithm2e}
\usepackage{amsmath}

\usepackage{amssymb}
\usepackage{bm}
\usepackage{enumitem}
\usepackage{soul}
\usepackage{threeparttable}
\usepackage{multirow}
\usepackage{makecell}
\usepackage[T1]{fontenc}
\usepackage{array}
\usepackage{listings}
\usepackage{xspace}
\usepackage{subcaption}
\usepackage{graphicx}
\usepackage{hyperref}
\usepackage{textcomp}
\usepackage{microtype}
\usepackage{calc}
\usepackage{pifont}
\usepackage{tcolorbox}
\usepackage{lscape}
\usepackage{rotating}
\usepackage{balance}
\usepackage{cite}

\newcolumntype{Y}{>{\centering\arraybackslash}X}

\SetCommentSty{mycommfont}
\SetKwInput{KwInput}{Input}
\SetKwInput{KwOutput}{Output}

\newlength{\ColorBoxDepthReference}
\newlength{\ColorBoxHeightReference}
\newlength{\Width}
\newcommand{\MyColorBox}[2][red]%
{%
	\settowidth{\Width}{#2}%
	\colorbox{#1}%
	{%
		\raisebox{-\ColorBoxDepthReference}%
		{%
			\parbox[b][\ColorBoxHeightReference+\ColorBoxDepthReference][c]{\Width}{\centering#2}%
		}%
	}%
}
\definecolor{codegreen}{rgb}{0,0.6,0}
\definecolor{codegray}{rgb}{0.5,0.5,0.5}
\definecolor{codepurple}{rgb}{0.58,0,0.82}
\usepackage{xcolor}

\usepackage{pgfplots}
\usepgfplotslibrary{groupplots}
\pgfplotsset{compat=1.18}

\begin{document}

\title{RAPO-Sol: Retrieval-Augmented Preference Optimization for Repository-Level Solidity Code Generation}

\author{
    \IEEEauthorblockN{
        Rongcun Wang\IEEEauthorrefmark{1}\thanks{Corresponding author.},
        Shi Chen\IEEEauthorrefmark{1},
        Yuan Tian\IEEEauthorrefmark{2},
        Rubing Huang\IEEEauthorrefmark{3}
    }

    \IEEEauthorblockA{
        \IEEEauthorrefmark{1}School of Computer Science and Technology (School of Artificial Intelligence),\\
        China University of Mining and Technology, Xuzhou, Jiangsu Province, China\\
        \IEEEauthorrefmark{2}Queen's University, Kingston, Canada\\
        \IEEEauthorrefmark{3}Macau University of Science and Technology, Macau, China\\
        Email: rcwang@cumt.edu.cn, ts24170102p31@cumt.edu.cn, y.tian@queensu.ca, rbhuang@must.edu.mo
    }
}

\maketitle

\begin{abstract}
Smart contracts, predominantly written in Solidity on Ethereum and other EVM-compatible blockchains, govern on-chain assets, permissions, and irreversible state changes. Writing them correctly is both labor-intensive and security-critical, making automated generation an appealing direction. The realistic target is not an isolated function but repository-level Solidity code generation: synthesizing a complete contract or library that a developer can ship directly. This task is challenging because a correct output requires cross-function consistency among state variables, modifiers, events, inheritance relations, external calls, and access-control logic. Moreover, small deviations in these constructs can change contract behavior even when the code remains syntactically plausible. To address these challenges, we present \textbf{RAPO-Sol}, a two-stage  framework for repository-level Solidity code generation. First, Retrieval-Augmented Fine-Tuning (RAFT) augments each training input with similar Solidity examples, allowing the model to learn recurring contract-level patterns while remaining retrieval-free at inference time. Second, Direct Preference Optimization (DPO) trains the model to prefer the reference contract over close but semantically flawed alternatives. To construct such rejected samples, we introduce Solidity Semantic-Anchor Perturbation (SAP), which perturbs validation statements, visibility modifiers, data-location keywords, context variables, payment operations, and low-level calls. We evaluate the framework on SolidityBench using CodeLlama-7B-Instruct, DeepSeek-Coder-6.7B-Instruct, and Qwen2.5-Coder-7B-Instruct. RAFT consistently outperforms standard supervised fine-tuning; SAP-based DPO further improves both BLEU and SolidityScore, a Solidity-aware embedding similarity metric, and the full RAFT+DPO pipeline achieves the best performance across all three models. These results indicate that retrieval during training and Solidity-aware preference data offer distinct, additive benefits for repository-level smart contract generation.
\end{abstract}

\section{Introduction}
\label{sec:introduction}

Smart contracts are a cornerstone of blockchain applications, especially in decentralized finance (DeFi). On Ethereum, smart contracts are most commonly written in Solidity, whose code defines application logic that directly manages on-chain assets, enforces access control, and triggers irreversible state transitions~\cite{Buterin2014Ethereum,Wood2014Ethereum}. Consequently, errors in Solidity code can have severe financial repercussions: mistakes in access control, handling of external calls, exception management, or state updates frequently lead to loss of funds or unsafe contract behavior~\cite{Atzei2017Survey,Luu2016Making}. Prior studies~\cite{Vidal2024OpenSCV,Durieux2020Empirical,Ferreira2020SmartBugs} further reveal that smart contract defects are predominantly semantic rather than merely syntactic. These characteristics make semantic correctness a central requirement for Solidity code generation, as even small semantic errors can compromise contract security or lead to financial loss.

Large language models (LLMs) have substantially advanced code generation. Code-oriented models such as CodeLlama~\cite{Roziere2023CodeLlama}, DeepSeek-Coder~\cite{Guo2024DeepSeekCoder}, and Qwen2.5-Coder~\cite{Hui2024Qwen25Coder} perform well on many programming tasks, and benchmarks such as HumanEval, CodeXGLUE, and MultiPL-E have driven progress in program synthesis and code understanding~\cite{Chen2021Evaluating,Lu2021CodeXGLUE,Cassano2023MultiPLE}. These benchmarks, however, primarily focus on single-function or single-file scenarios in general-purpose languages. They reveal little about whether a model can produce a complete, internally consistent smart contract, where long-range dependencies and Solidity-specific semantic constraints make correctness a whole-contract property.
In this paper, we study \textit{repository-level Solidity generation at the contract-unit granularity}: the task is to generate a complete contract or library unit from a natural-language requirement, covering imports, inheritance relations, interfaces, library usage, state declarations, and cross-function logic, rather than completing an isolated function. A correct output requires consistency across state variables, modifiers, events, inheritance, access checks, and cross-function behavior. This need for long-range consistency aligns with findings from repository-level code generation, where project-level context remains difficult for current models~\cite{Shrivastava2023Repository,Zhang2023RepoCoder,Ding2023CrossCodeEval}.

Recent Solidity benchmarks consider more realistic development settings. SolEval~\cite{Peng2025SolEval}, for example, evaluates repository-level Solidity generation by prompting the model with a function signature, a natural-language requirement, and surrounding repository context. The generated function is then injected back into the original repository for execution-based scoring using $\text{Pass}@k$. Because the contract structure, including imports, inheritance, interfaces, and state declarations, is already fixed by the surrounding scaffold, this setting evaluates local function synthesis rather than structural synthesis itself. In contrast, generating a complete contract or library directly from a short requirement raises two coupled challenges: (1) inferring a coherent, multi-component contract structure, and (2) producing Solidity code that is not only syntactically plausible but semantically correct in its handling of access control, data locations, and payment-sensitive operations. SolidityBench~\cite{Chen2026RepositoryLevelSolidity} directly matches this setting, and we therefore adopt it as our evaluation benchmark.

For this task,  supervised fine-tuning (SFT) achieves promising results \cite{Chen2026RepositoryLevelSolidity}. On its own, however, SFT faces two primary limitations in this setting, each motivating a distinct form of supplementary supervision. First, SFT imitates reference contracts, but a natural-language requirement rarely specifies the full contract structure. Although such patterns may exist in the training corpus, standard SFT does not explicitly condition each training instance on similar contracts that capture reusable contract-level structures. Retrieval-augmented generation (RAG) can surface such patterns during inference~\cite{Lewis2020Retrieval}, yet this ties inference to retriever quality and lengthens the prompt. Retrieval-Augmented Fine-Tuning (RAFT) instead supplies retrieved examples during training only, so the model learns from related contracts while remaining retrieval-free during inference~\cite{Zhang2024RAFT}. Second, SFT offers only positive demonstrations and never contrasts a correct contract with a nearby variant that appears syntactically plausible but behaves differently. In Solidity, such near misses are common: a one-line change to an access-control guard, a visibility modifier, a data-location keyword, or a payment-sensitive call can preserve syntactic plausibility while silently changing contract semantics. Preference optimization can provide this contrast, and Direct Preference Optimization (DPO) learns directly from chosen and rejected responses without training a separate reward model~\cite{Rafailov2023Direct}. The central design challenge is not merely applying DPO, but constructing rejected contracts that stay close to the reference while being wrong in Solidity-specific ways.

To provide the complementary supervision required by these two limitations, we propose \textbf{RAPO-Sol} (\textbf{R}etrieval-\textbf{A}ugmented \textbf{P}reference \textbf{O}ptimization for Repository-Level \textbf{Sol}idity Code Generation)\footnote{\url{https://github.com/ChenS0827/RAPO-Sol}}, a novel two-stage  framework. In the first stage, RAFT incorporates retrieved requirement–code examples during fine-tuning, enabling the model to learn reusable contract-level structures while retaining requirement-only inference. In the second stage, DPO refines the RAFT-trained model by contrasting each reference contract with a structurally close but semantically flawed alternative. These rejected contracts are constructed using our Solidity Semantic-Anchor Perturbation (SAP), which perturbs condition, visibility, data/context, and call/payment anchors.

We evaluate RAPO-Sol on SolidityBench~\cite{Chen2026RepositoryLevelSolidity}, a dataset of 5,470 repository-level requirement–contract pairs, using CodeLlama-7B-Instruct-HF~\cite{Roziere2023CodeLlama}, DeepSeek-Coder-6.7B-Instruct~\cite{Guo2024DeepSeekCoder}, and Qwen2.5-Coder-7B-Instruct~\cite{Hui2024Qwen25Coder}. We report BLEU~\cite{Papineni2002BLEU} and SolidityScore~\cite{Chen2026RepositoryLevelSolidity}, where SolidityScore is a Solidity-aware embedding-based similarity metric that complements surface-level overlap by measuring contextual alignment between generated and reference contracts. Experimental results show that RAFT consistently outperforms standard SFT, and SAP-based DPO brings further improvements across all three models.  RAPO-Sol achieves the best results on both metrics. For example, on Qwen2.5-Coder, RAPO-Sol improves BLEU from 35.962 to 41.263 and SolidityScore from 0.653 to 0.707 over SFT. These results suggest that retrieval during training and SAP-based preference data provide complementary supervision for complete-contract Solidity generation.

Overall, the paper makes the following contributions:

\begin{itemize}
    \item We propose \textbf{RAPO-Sol}, a two-stage framework that integrates RAFT with DPO while keeping inference retrieval-free.

    \item We introduce \textbf{Solidity SAP}, a domain-aware hard-negative construction strategy that perturbs condition, visibility, data/context, and call/payment anchors to create close but semantically flawed rejected contracts, enabling DPO to learn fine-grained Solidity semantics.

    \item  We conduct a systematic empirical study on SolidityBench, a dataset containing 5,470 requirement-contract pairs across three code models, demonstrating complementary gains from the two-stage pipeline and isolating the effects of each component through controlled ablations.
\end{itemize}

The remainder of this paper is organized as follows. Section~\ref{sec:motivation} describes the research motivation. Section~\ref{sec:method} presents the proposed framework. The experimental setup and results are detailed in Section~\ref{sec:evaluation_setup} and Section~\ref{sec:results}, respectively. Section~\ref{sec:threats} discusses the threats to validity. Section~\ref{sec:related_work} summarizes related work, followed by the conclusion in Section~\ref{sec:conclusion}.

\section{Motivation}
\label{sec:motivation}

Repository-level Solidity code generation introduces two primary challenges that expose the limitations of existing code generation models. First, the model needs to infer a global contract structure that is often implicit in the requirements, encompassing state variables, modifiers, events, inheritance, interface calls, helper functions, and cross-function updates. This structural context is frequently scattered across files, parent contracts, and project dependencies, making it difficult for current models to assemble the long-range dependencies needed for a coherent contract~\cite{Shrivastava2023Repository,Zhang2023RepoCoder,Ding2023CrossCodeEval}. 
Second, the model needs to correctly handle fine-grained Solidity details where a small textual footprint can have a disproportionate behavioral effect. Omitting a validation check, changing a visibility modifier, using an incorrect data location, or altering payment behavior can leave a contract syntactically plausible while silently changing authorization, data flow, or value-transfer semantics~\cite{Atzei2017Survey,Luu2016Making,Tsankov2018Securify,Durieux2020Empirical}. 

These two challenges are distinct but interact: the structural patterns that coordinate state and logic are precisely the points where fine-grained omissions produce critical vulnerabilities. This interplay has direct consequences for training: standard SFT, which maximizes the likelihood of reference contracts, does not explicitly penalize functionally incorrect variants. Fig.~\ref{fig1_motivating_example} illustrates such a case: a one-line edit removes an access-control check while leaving the contract nearly unchanged. Because SFT optimizes token-level likelihood without distinguishing security-critical tokens, a model may achieve low perplexity while still missing a few checks that are essential to contract semantics. Visibility modifiers, data locations, and payment-sensitive constructs share the same characteristic: small textual changes can profoundly affect authorization, interface behavior, gas consumption, or value transfer. SFT provides no targeted signal to disambiguate them.

These two challenges call for complementary mechanisms that respectively address structural coherence and semantic precision. For the structural challenge, retrieval offers a natural signal: similar contracts often share common blueprints involving ownership checks, event emission, guarded updates, and external-call patterns. However, inference-time retrieval~\cite{Lewis2020Retrieval} introduces a retriever dependency and longer prompts at deployment. Since inference uses only the requirement, we adopt RAFT~\cite{Zhang2024RAFT}, which uses retrieved examples during fine-tuning to teach the model structural patterns while leaving the final inference input as the requirement alone. 

\lstdefinelanguage{Solidity}{
  morekeywords={pragma,solidity,import,contract,is,function,external,public,
    internal,private,payable,nonReentrant,require,return,returns,mapping,
    address,uint256,bool,memory,calldata,storage,msg,emit,modifier},
  sensitive=true,
  morecomment=[l]{//},
  morestring=[b]",
}

\lstdefinestyle{solstyle}{
  language=Solidity,
  basicstyle=\scriptsize\ttfamily,
  keywordstyle=\color{codepurple},
  commentstyle=\color{codegray}\itshape,
  stringstyle=\color{codegreen},
  breaklines=true,
  columns=fullflexible,
  keepspaces=true,
  aboveskip=2pt,
  belowskip=2pt,
  frame=none,
  escapeinside={(*@}{@*)},
}

\begin{figure}[!htbp]
\centering
\begin{tcolorbox}[
  colback=gray!4!white,
  colframe=gray!70!black,
  width=\linewidth,
  arc=1mm,
  boxrule=0.6pt,
  left=3pt,right=3pt,top=2pt,bottom=2pt,
  title=\textbf{\small A close but semantically flawed Solidity variant}
]
\scriptsize
\textbf{Requirement.} \textit{Let only the owner withdraw the full
contract balance to a recipient address.}

\smallskip
\textbf{Reference contract $y_w$ (chosen).}
\begin{lstlisting}[style=solstyle]
address public owner;
function withdraw(address to) external {
  (*@\colorbox{green!12}{\texttt{require(msg.sender == owner, "not owner");}}@*)
  (bool ok,) = to.call{value: address(this).balance}("");
  require(ok, "transfer failed");
}
\end{lstlisting}

\smallskip
\textbf{SAP-perturbed contract $y_l$ (rejected).}
\begin{lstlisting}[style=solstyle]
address public owner;
function withdraw(address to) external {
  (*@\colorbox{red!12}{\texttt{// owner check removed by SAP}}@*)
  (bool ok,) = to.call{value: address(this).balance}("");
  require(ok, "transfer failed");
}
\end{lstlisting}
\end{tcolorbox}
\vspace{-0.3em}
\caption{A close Solidity variant with a flawed condition anchor. Removing the \texttt{require} statement that enforces the owner check keeps the contract syntactically valid and highly similar to the reference contract $y_w$, but allows any caller to withdraw the contract balance.}
\label{fig1_motivating_example}
\end{figure}

For the semantic challenge, we turn to preference optimization, which explicitly contrasts correct and incorrect implementations. We use DPO~\cite{Rafailov2023Direct} because it learns directly from chosen and rejected pairs without training a separate reward model. The key is constructing rejected samples that are close to the reference in surface form but wrong along Solidity-specific dimensions. Superficial edits may be too weak, while heavily corrupted programs are trivially distinguishable. We therefore design Solidity SAP, a perturbation strategy that targets condition, visibility, data/context, and call/payment anchors, the very constructs whose omission or alteration leads to the vulnerabilities described above.  

Together, these observations motivate our two-stage design: RAFT addresses the structural challenge by exposing the model to related contracts during training, while SAP-based DPO tackles the semantic challenge by teaching the model to prefer chosen contracts over Solidity-specific near-miss variants. The goal is to strengthen the training signal for repository-level Solidity generation, not to replace compiler checks, testing, or security analysis.
\section{Approach}
\label{sec:method}

\subsection{Overview}
\label{subsec:framework}

Given a training set $\mathcal{D}_{\mathrm{train}}=\{(x_i,y_i)\}_{i=1}^{n}$, where $x_i$ is a natural-language requirement and $y_i$ is the reference Solidity contract or contract library, our goal is to train a model that generates a contract $\hat{y}_i$ from $x_i$. As discussed in Section~\ref{sec:motivation}, this task requires both reusable contract-level structure and fine-grained Solidity semantic precision, the ability to distinguish correct implementations from near-miss variants with subtle semantic errors. RAPO-Sol addresses these two requirements through a two-stage training framework, as illustrated in Fig.~\ref{fig2_framework_overview}.

\begin{figure*}[!htbp]
\centering
\resizebox{0.85\textwidth}{!}{
\begin{tikzpicture}[
    font=\small,
    >=Latex,
    line/.style={-Latex, line width=0.8pt, draw=black!65},
    redline/.style={-Latex, line width=0.95pt, draw=red!72!black},
    grayline/.style={-Latex, line width=0.75pt, draw=black!42},
    plainred/.style={line width=0.75pt, draw=red!70!black},
    stagebox/.style={
        rounded corners=17pt,
        fill=black!2.5,
        draw=black!10,
        line width=0.55pt
    },
    title/.style={
        font=\large\bfseries,
        align=center
    },
    box/.style={
        rounded corners=6pt,
        line width=0.9pt,
        minimum height=0.88cm,
        align=center,
        inner xsep=6pt,
        inner ysep=5pt
    },
    greenbox/.style={
        box,
        draw=green!45!black,
        fill=green!11
    },
    bluebox/.style={
        box,
        draw=blue!60!black,
        fill=blue!9
    },
    redbox/.style={
        box,
        draw=red!65!black,
        fill=red!7
    },
    orangebox/.style={
        box,
        draw=orange!80!black,
        fill=orange!13
    },
    modelbox/.style={
        box,
        draw=blue!65!black,
        fill=blue!10,
        minimum width=1.75cm
    },
    modelred/.style={
        box,
        draw=red!65!black,
        fill=red!6,
        minimum width=1.85cm
    },
    strategy/.style={
        rounded corners=5pt,
        draw=orange!80!black,
        fill=orange!8,
        line width=0.75pt,
        align=center,
        font=\scriptsize,
        minimum width=2.25cm,
        minimum height=0.43cm,
        inner xsep=5pt,
        inner ysep=2.5pt
    },
    strategyours/.style={
        strategy,
        draw=red!70!black,
        fill=red!7,
        font=\scriptsize\bfseries,
        minimum width=2.45cm
    },
    bridge/.style={
        rounded corners=5pt,
        draw=black!24,
        fill=white,
        align=center,
        font=\scriptsize,
        inner xsep=8pt,
        inner ysep=3.5pt
    }
]

\def\stageCenterX{1.50}
\def\stageWidth{23.40cm}

\begin{scope}[on background layer]
    \node[stagebox, minimum width=\stageWidth, minimum height=3.00cm]
        (stageonebg) at (\stageCenterX, 2.00) {};
    \node[stagebox, minimum width=\stageWidth, minimum height=3.55cm]
        (stagetwobg) at (\stageCenterX, -1.55) {};
\end{scope}

\node[title] at (\stageCenterX, 3.08)
{STAGE 1: Retrieval-Augmented Fine-Tuning (RAFT)};

\node[greenbox, minimum width=1.70cm] (input) at (-4.85, 1.95)
{\textbf{Input}\\requirement\\[-1pt]$x_i$};

\node[bluebox, minimum width=1.95cm] (retriever) at (-1.90, 1.95)
{\textbf{Retriever}\\[-1pt]$\mathcal{R}(x_i,\mathcal{D}_{\mathrm{train}})$};

\node[bluebox, minimum width=2.80cm] (topk) at (1.50, 1.95)
{\textbf{Top-$k$ examples}\\[-1pt]
$\mathcal{E}_i=\{(x_{j_1},y_{j_1}),\ldots\}$};

\node[greenbox, minimum width=2.30cm] (aug) at (5.10, 1.95)
{\textbf{Augmented input}\\[-1pt]$\tilde{x}_i=[x_i;\mathcal{E}_i]$};

\node[modelbox, minimum width=1.95cm] (raft) at (8.20, 1.95)
{\textbf{RAFT policy}\\[-1pt]$\pi_{\mathrm{RAFT}}$};

\draw[line] (input.east) -- (retriever.west);
\draw[line] (retriever.east) -- (topk.west);
\draw[line] (topk.east) -- (aug.west);
\draw[redline] (aug.east) -- (raft.west);

\node[bridge] (bridge) at (8.20, 0.50)
{initialize DPO\\from $\pi_{\mathrm{RAFT}}$};

\draw[grayline] (raft.south) -- (bridge.north);

\node[title] at (\stageCenterX, -0.16)
{STAGE 2: Direct Preference Optimization (DPO)};

\node[greenbox, minimum width=1.95cm, minimum height=0.82cm] (sample) at (-8.05, -1.20)
{\textbf{Training sample}\\[-1pt]$(x_i,y_i)$};

\node[greenbox, minimum width=1.95cm, minimum height=0.82cm] (chosen) at (-4.90, -1.20)
{\textbf{Chosen}\\reference code\\[-1pt]$y_w \leftarrow y_i$};

\node[redbox, minimum width=2.15cm, minimum height=0.82cm] (perturb) at (-1.65, -1.20)
{\textbf{Rejected-response}\\construction};

\node[orangebox, minimum width=1.95cm, minimum height=0.82cm] (rejected) at (1.60, -1.20)
{\textbf{Rejected}\\hard negative\\[-1pt]$y_l$};

\node[redbox, minimum width=2.15cm, minimum height=0.82cm] (triple) at (4.85, -1.20)
{\textbf{Preference triple}\\[-1pt]$(x_i,y_w,y_l)$};

\node[modelbox, minimum width=2.25cm, minimum height=0.82cm] (initpolicy) at (8.20, -1.20)
{\textbf{Initial policy}\\[-1pt]$\pi_{\theta}\leftarrow\pi_{\mathrm{RAFT}}$};

\node[modelred, minimum width=2.00cm, minimum height=0.82cm] (ours) at (11.35, -1.20)
{\textbf{Final policy}\\[-1pt]$\pi_{\mathrm{RAPO}}$};

\draw[line] (sample.east) -- (chosen.west);
\draw[redline] (chosen.east) -- (perturb.west);
\draw[redline] (perturb.east) -- (rejected.west);
\draw[line] (rejected.east) -- (triple.west);
\draw[line] (triple.east) -- (initpolicy.west);
\draw[redline] (initpolicy.east) -- (ours.west);

\draw[grayline] (bridge.south) -- (initpolicy.north);

\node[
    font=\scriptsize\bfseries,
    text=red!70!black,
    align=center
] (strategytitle) at (-1.65, -2.05)
{rejected-response construction};

\node[strategy, minimum width=0.82cm, inner xsep=1.5pt] (surface) at (-3.55, -2.90)
{SLP};

\node[strategy, minimum width=0.82cm, inner xsep=1.5pt] (logic) at (-1.65, -2.90)
{GLP};

\node[strategyours, minimum width=0.82cm, inner xsep=1.5pt] (semantic) at (0.25, -2.90)
{SAP};

\coordinate (branchMid) at (-1.65, -2.52);
\coordinate (branchL) at (surface.north |- branchMid);
\coordinate (branchM) at (logic.north |- branchMid);
\coordinate (branchR) at (semantic.north |- branchMid);

\draw[plainred] (perturb.south) -- (strategytitle.north);

\draw[plainred] (strategytitle.south) -- (branchM);

\draw[plainred] (branchL) -- (branchR);

\draw[plainred] (branchL) -- (surface.north);
\draw[plainred] (branchM) -- (logic.north);
\draw[plainred] (branchR) -- (semantic.north);

\end{tikzpicture}
}
\caption{Overview of RAPO-Sol. Stage~1 uses retrieved requirement-code examples only during fine-tuning to learn reusable contract-level structures. Stage~2 initializes from the RAFT policy and applies DPO on preference triples, where the reference contract is the chosen response and a Solidity-aware perturbed variant is the rejected response.}
\label{fig2_framework_overview}
\end{figure*}

Stage~1 trains a base policy with RAFT. Each training requirement is augmented with retrieved requirement-code examples from the training set, while the prediction target remains the current reference contract. These examples expose the model to recurring Solidity structures, such as ownership control, guarded state updates, event emission, and external-call logic. Retrieval is used only during fine-tuning; at inference time, the model receives only the natural-language requirement.

Stage~2 refines the RAFT policy with DPO~\cite{Rafailov2023Direct}. Each preference triple is constructed from the same training sample: the reference contract is used as the chosen response, and a close but semantically flawed variant is used as the rejected response. This contrastive signal encourages the model to preserve Solidity-critical details that likelihood-based fine-tuning may underemphasize, including validation checks, visibility boundaries, data/context usage, and payment-related operations. Since this stage also uses requirement-only inputs, the final RAPO-Sol remains retrieval-free at inference time.

\subsection{Retrieval-Augmented Fine-Tuning}
\label{subsec:raft}

Standard SFT trains the model on requirement-contract pairs by maximizing the likelihood of the reference contract. In repository-level Solidity generation, however, the requirement often does not explicitly specify all state variables, modifiers, events, inherited components, or helper functions. The model must therefore infer much of the contract-level structure from limited supervision, which can lead to locally plausible code that misses required global structure~\cite{Shrivastava2023Repository,Zhang2023RepoCoder,Peng2025SolEval}. Moreover, because SFT trains on each requirement-contract pair in isolation, it cannot explicitly use similar contracts as context for learning reusable Solidity structures.  Consequently, a model trained this way may produce defects closely aligned with those documented in prior studies~\cite{Luu2016Making,Durieux2020Empirical}.

RAFT addresses this limitation by introducing related examples during fine-tuning. Before training, we index all requirements from the training split using BM25. For each training sample, we query this index with the current requirement and prepend the top-$k$ matched requirement-code pairs to the model input. These examples serve as contextual cues that expose the model to recurring Solidity patterns, such as ownership control, event emission, mapping updates,  and guarded external calls. Crucially, the prediction target remains the current reference contract; the retrieved examples function purely as conditioning context, not as templates to be copied. At inference time, retrieval is removed and the model receives only the requirement.

Formally, given a training sample $(x_i, y_i)$, where $x_i$ is the natural-language requirement and $y_i$ is the reference Solidity code, we retrieve $k$ examples from the training set $\mathcal{D}_{\mathrm{train}}$. The function $\operatorname{BM25}(x_i, x_j)$ denotes the lexical relevance score between the current requirement and a candidate requirement. The current sample is excluded by the condition $j \neq i$, and $(x_{j_m}, y_{j_m})$ denotes the $m$-th selected requirement-code pair:
\begin{equation}
\label{eq:retrieve}
\begin{aligned}
\mathcal{E}_i
&= \operatorname{top-k}_{(x_j,y_j) \in \mathcal{D}_{\mathrm{train}}, \, j \neq i}
\; \operatorname{BM25}(x_i, x_j) \\[2pt]
&= \{(x_{j_1}, y_{j_1}), \ldots, (x_{j_k}, y_{j_k})\}.
\end{aligned}
\end{equation}

The retrieved examples are concatenated with the current requirement to form the augmented input:
\begin{equation}
\tilde{x}_i = \operatorname{Concat}(x_i, \mathcal{E}_i).
\label{eq:aug_input}
\end{equation}
Here, $\operatorname{Concat}(\cdot)$ keeps the current requirement as the leading task description and appends retrieved examples as requirement-code blocks ordered by BM25 score, using fields such as \texttt{Requirement}, \texttt{Example Requirement}, and \texttt{Example Code}.

The model is then optimized with the standard autoregressive language modeling objective, where $\pi_\theta$ denotes the trainable model, $y_{i,t}$ is the $t$-th target token, and $y_{i,<t}$ is the target prefix under teacher forcing:
\begin{equation}
\mathcal{L}_{\mathrm{RAFT}} = - \sum_{(x_i, y_i) \in \mathcal{D}_{\mathrm{train}}}
\sum_{t=1}^{|y_i|}
\log \pi_{\theta}(y_{i,t} \mid \tilde{x}_i, y_{i,<t}).
\label{eq:raft_loss}
\end{equation}

Compared with standard SFT, RAFT modifies only the training input while leaving the supervised objective unchanged~\cite{Zhang2024RAFT}. This design is critical for deployment: the final model requires no retriever, no retrieval index, and no extended prompts at inference time. In other words, RAFT uses retrieval to enrich training context, but the optimization target remains the original reference contract.

\subsection{Preference Dataset Construction}
\label{subsec:preference_data}

RAFT improves exposure to contract-level patterns, but it still optimizes a likelihood objective. It does not explicitly compare a correct contract with a nearby contract that violates a Solidity-specific constraint. We therefore build a preference dataset for DPO. Each sample is a triplet $(x_i, y_w, y_l)$, where $x_i$ is the requirement, $y_w$ is the reference contract used as the chosen response, and $y_l$ is a perturbed version of the same contract used as the rejected response.

The rejected response needs to satisfy two conditions. First, it should remain close enough to the reference to serve as a nontrivial hard negative; otherwise, DPO may learn only a shallow distinction. Second, the perturbation should affect behavior that matters for Solidity. Changes that only modify formatting or comments are unlikely to teach the model to preserve contract semantics. The three strategies below instantiate this design space.

\subsubsection{Surface-Level Perturbation (SLP)}
\label{subsubsec:surface_perturbation}

Surface-level perturbation changes the appearance of the code while largely preserving behavior. We rename variables, add unused statements, insert irrelevant comments, or make minor textual edits. Similar transformations are often used in code augmentation and robustness studies~\cite{Yu2022Data,Yefet2020Adversarial,Na2023DIP}. We include this strategy as an intentionally weak, low-semantic baseline: because many such edits preserve behavior, it tests whether DPO gains can be explained by superficial contrast alone.

\subsubsection{General Logical Perturbation (GLP)}
\label{subsubsec:logical_perturbation}

General logical perturbation changes local program behavior in a language-agnostic way. We flip comparison operators, replace arithmetic operators, or swap nearby statements. These edits follow the spirit of mutation testing and recent mutation-based evaluation of code LLMs~\cite{Jia2011Analysis,Li2024Mutation}. This strategy creates more challenging rejected samples than surface-level perturbation, but it does not target Solidity-specific boundaries such as data locations, caller context, value transfer, or function visibility. We apply the same structural validity filter used by SAP to ensure rejected samples remain usable as preference responses.

\subsubsection{Solidity Semantic-Anchor Perturbation}
\label{subsubsec:sap}

Solidity SAP is our main strategy for constructing Solidity-aware hard negatives.
It targets Solidity constructs whose local textual changes can substantially affect contract behavior. According to previous studies~\cite{Atzei2017Survey,Luu2016Making,Vidal2024OpenSCV,Durieux2020Empirical,Tikhomirov2018SmartCheck}, security-critical defects in smart contracts often concentrate in a few key areas: validation and access-control logic, interface exposure, execution-context and data handling, and value-transfer or external-call operations~\cite{SolidityDocs}. We therefore map these defect-prone areas to four concrete anchor types for perturbation: condition, visibility, data/context, and call/payment anchors. Specifically, condition anchors cover validation-related constructs such as \texttt{require}, \texttt{assert}, and guard expressions like \texttt{if (msg.sender != owner) revert}. Visibility anchors target Solidity visibility keywords, including \texttt{public}, \texttt{private}, \texttt{internal}, and \texttt{external}. Data/context anchors encompass data location keywords and execution-context variables, such as \texttt{calldata}, \texttt{memory}, \texttt{storage}, \texttt{msg.sender}, and \texttt{msg.value}. Call/payment anchors cover payable-related operations and low-level call patterns.

Table~\ref{table1_sap_operations} presents the anchor types and representative perturbation operations. The edits are intentionally local. For condition anchors, SAP removes or weakens validation guards such as \texttt{require} and \texttt{assert}. For visibility anchors, it changes function visibility or state-variable exposure. For data/context anchors, it perturbs data location keywords or context variables such as \texttt{msg.sender}. For call/payment anchors, it alters payable annotations, value-transfer operations, or low-level call handling. Although these edits affect only a small part of the code, they can change authorization, data flow, value transfer, or external interaction behavior~\cite{Jiang2018ContractFuzzer,Feist2019Slither,Ferreira2020SmartBugs}.

\begin{table}[!htbp]
\centering
\caption{SAP anchor types, representative perturbation operations, and examples}
\label{table1_sap_operations}
\tiny
\setlength{\tabcolsep}{3.2pt}
\renewcommand{\arraystretch}{1.20}
\begin{tabularx}{\linewidth}{
@{}
>{\raggedright\arraybackslash}p{0.16\linewidth}
>{\raggedright\arraybackslash}p{0.23\linewidth}
>{\raggedright\arraybackslash}X
@{}}
\toprule
\textbf{Anchor type} &
\textbf{Representative constructs} &
\textbf{Perturbation operations and examples} \\
\midrule

Condition &
\makecell[l]{\texttt{require}, \texttt{assert},\\
guard expressions} &
\makecell[l]{Remove or weaken validation guards; alter guard conditions.\\
\textit{Example:} \texttt{require(msg.sender == owner)} $\rightarrow$ removed;\\
\texttt{a \&\& b} $\rightarrow$ \texttt{a || b};\\
\texttt{amount <= balance} $\rightarrow$ \texttt{amount < balance}.} \\

\addlinespace[2pt]

Visibility &
\makecell[l]{\texttt{public}, \texttt{private},\\
\texttt{internal}, \texttt{external}} &
\makecell[l]{Change function visibility or state-variable exposure.\\
\textit{Example:} \texttt{internal} $\rightarrow$ \texttt{public};\\
\texttt{private} $\rightarrow$ \texttt{public}.} \\

\addlinespace[2pt]

Data/context &
\makecell[l]{data-location\\
keywords and\\
context variables} &
\makecell[l]{Swap data-location keywords; replace context variables.\\
\textit{Example:} \texttt{calldata} $\rightarrow$ \texttt{memory};\\
\texttt{msg.sender} $\rightarrow$ \texttt{tx.origin};\\
\texttt{msg.value} $\rightarrow$ \texttt{0}.} \\

\addlinespace[2pt]

Call/payment &
\makecell[l]{\texttt{payable}, \texttt{transfer},\\
\texttt{send}, low-level\\
\texttt{call}} &
\makecell[l]{Add/remove payable annotations, value transfer, or call handling.\\
\textit{Example:} alter \texttt{payable};\\
\texttt{transfer(...)} $\rightarrow$ \texttt{send(...)};\\
remove \texttt{require(ok)} after low-level \texttt{call}.} \\

\bottomrule
\end{tabularx}
\end{table}

Algorithm~\ref{alg:sap} shows the process for constructing SAP hard negatives. Given a requirement $x_i$, chosen code $y_w$, seed set $S$, anchor set $\mathcal{A}$, and perturbation ratio $\tau$, SAP first locates all anchor occurrences in $y_w$. Let $\mathcal{L}$ denote the resulting anchor list, and let $N = |\mathcal{L}|$ be the number of detected anchors. If no anchor is found, SAP skips the sample rather than forcing an unrelated edit. Otherwise, it sets $m=\min(N,\max(1,\lfloor \tau \cdot N \rfloor))$,
ensuring that at least one anchor is perturbed while never selecting more anchors than are available. Then, for each seed, SAP copies the chosen code $y_w$ to initialize the rejected code $y_l$, samples $m$ anchors from $\mathcal{L}$, and applies the corresponding rule-based perturbation operations to $y_l$. Finally, the resulting candidate is retained only if it differs from $y_w$ and passes
\textsc{IsValid}; the surviving triplet $(x_i, y_w, y_l)$ is added to the preference set $\mathcal{P}$.

\textsc{IsValid} is a lightweight structural filter we developed to check whether a perturbation produces a usable rejected response. It is not intended as a proof of correctness. It discards candidates that are unchanged, empty, lose the contract or library declaration, remove the target body, or contain obvious delimiter imbalance after perturbation. This filter keeps rejected responses structurally valid for preference optimization, but it does not certify compilation, functional correctness, or security.

\begin{algorithm}[!htbp]
\captionsetup{justification=centering,singlelinecheck=false}
\SetAlCapFnt{\small}
\SetAlCapNameFnt{\small}
\caption{Solidity Semantic-Anchor Perturbation (SAP)}
\label{alg:sap}
\small
\setlength{\tabcolsep}{0pt}
\renewcommand{\arraystretch}{0.88}
\begin{tabular}{@{}r@{\hspace{0.35em}}p{0.82\linewidth}@{}}
    & \textbf{Input:} Requirement $x_i$, chosen code $y_w$, seed set $S$ with $|S|=s$, perturbation ratio $\tau$; anchor set $\mathcal{A}$ with condition ($\mathcal{A}_{\mathrm{cond}}$), visibility ($\mathcal{A}_{\mathrm{vis}}$), data/context ($\mathcal{A}_{\mathrm{ctx}}$), and call/payment ($\mathcal{A}_{\mathrm{call}}$) anchors; \\
    & \textbf{Output:} Preference triples $\mathcal{P}=\{(x_i,y_w,y_l)\}$; \\

1:  & $\mathcal{P} \leftarrow \emptyset$; \\
2:  & \textcolor{gray}{// Locate Solidity semantic anchors in the chosen contract} \\
3:  & $\mathcal{L} \leftarrow \textsc{LocateAnchors}(y_w,\mathcal{A})$; \\
4:  & $N \leftarrow |\mathcal{L}|$; \\
5:  & \textbf{if} $N=0$ \textbf{then return} $\mathcal{P}$; \\
6:  & $m \leftarrow \min\bigl(N,\max(1,\lfloor \tau \cdot N \rfloor)\bigr)$; \\

7:  & \textbf{for each} seed $r$ in $S$ \textbf{do} \\
8:  & \quad $y_l \leftarrow \textsc{Copy}(y_w)$; \\
9:  & \quad \textcolor{gray}{// Select anchors and apply type-specific perturbations} \\
10: & \quad $\mathcal{T} \leftarrow \textsc{RandomSelect}(\mathcal{L},m,r)$; \\
11: & \quad \textbf{for each} anchor $a$ in $\mathcal{T}$ \textbf{do} \\
12: & \quad\quad \textbf{if} $a \in \mathcal{A}_{\mathrm{cond}}$ \textbf{then} \\
13: & \quad\quad\quad $y_l \leftarrow \textsc{PerturbCondition}(y_l,a)$; \\
14: & \quad\quad \textbf{else if} $a \in \mathcal{A}_{\mathrm{vis}}$ \textbf{then} \\
15: & \quad\quad\quad $y_l \leftarrow \textsc{PerturbVisibility}(y_l,a)$; \\
16: & \quad\quad \textbf{else if} $a \in \mathcal{A}_{\mathrm{ctx}}$ \textbf{then} \\
17: & \quad\quad\quad $y_l \leftarrow \textsc{PerturbDataContext}(y_l,a)$; \\
18: & \quad\quad \textbf{else if} $a \in \mathcal{A}_{\mathrm{call}}$ \textbf{then} \\
19: & \quad\quad\quad $y_l \leftarrow \textsc{PerturbCallPayment}(y_l,a)$; \\
20: & \quad\quad \textbf{end if} \\
21: & \quad \textbf{end for} \\

22: & \quad \textcolor{gray}{// Retain only changed and structurally valid variants} \\
23: & \quad \textbf{if} $y_l\neq y_w$ \textbf{and} \textsc{IsValid}$(y_l)$ \textbf{then} \\
24: & \quad\quad $\mathcal{P} \leftarrow \mathcal{P} \cup \{(x_i,y_w,y_l)\}$; \\
25: & \quad \textbf{end if} \\
26: & \textbf{end for} \\
27: & \textbf{return} $\mathcal{P}$; \\
\end{tabular}
\end{algorithm}

The output is a set of preference triples $\mathcal{P}=\{(x_i,y_w,y_l)\}$. The chosen contract $y_w$ and rejected response $y_l$ remain close in surface form, while $y_l$ contains a targeted SAP perturbation. This provides DPO with a direct comparison between preserving and violating a Solidity-relevant semantic anchor.

\subsection{Direct Preference Optimization}
\label{subsec:dpo}
 Given the constructed preference dataset, we further optimize the RAFT policy using DPO~\cite{Rafailov2023Direct}. DPO increases the relative log-likelihood of the chosen response over the rejected response while constraining the updated policy with a frozen reference model. This fits our setting because many  repository-level Solidity samples depend on imported contracts, library paths, compiler versions, or deployment assumptions that are not fully available in a standalone benchmark instance~\cite{Peng2025SolEval}. Without a complete execution environment, methods that require online feedback (e.g., compiler diagnostics or test results) are impractical. Offline preference optimization therefore provides a controlled training signal without requiring every sample to be reconstructed as an executable project. In the RAFT+DPO setting, the frozen reference model $\pi_{\mathrm{ref}}$ is a copy of the RAFT checkpoint used to initialize the trainable policy $\pi_\theta$; for the SFT+DPO ablation, $\pi_{\mathrm{ref}}$ is analogously copied from the SFT checkpoint.

For an input requirement $x_i$, chosen code $y_w$, and rejected code $y_l$, DPO optimizes the following loss over the preference dataset $\mathcal{D}_{\mathrm{pref}}$:

\begin{equation}
\small
\begin{aligned}
\mathcal{L}_{\mathrm{DPO}}
:= -\mathbb{E}_{(x_i,y_w,y_l)\sim \mathcal{D}_{\mathrm{pref}}}
\bigl[
\log \sigma\bigl(\beta \Delta(x_i,y_w,y_l)\bigr)
\bigr].
\end{aligned}
\label{eq:dpo_loss}
\end{equation}

\begin{equation}
\small
\begin{aligned}
\Delta(x_i,y_w,y_l)
&=
\log
\frac{\pi_\theta(y_w\mid x_i)}
{\pi_{\mathrm{ref}}(y_w\mid x_i)}
-
\log
\frac{\pi_\theta(y_l\mid x_i)}
{\pi_{\mathrm{ref}}(y_l\mid x_i)} .
\end{aligned}
\label{eq:dpo_delta}
\end{equation}

In Eq.~(\ref{eq:dpo_loss}), $\mathcal{D}_{\mathrm{pref}}$ contains the SAP or baseline preference triples, $\pi_{\theta}$ is the trainable policy model, $\pi_{\mathrm{ref}}$ is the frozen reference model used to constrain updates, $\beta$ controls the strength of the preference constraint, and  $\sigma(\cdot)$ is the sigmoid function. The terms $\log \pi_\theta(y \mid x_i)$ and $\log \pi_{\mathrm{ref}}(y \mid x_i)$ denote sequence log-probabilities, computed as the sum of token log-probabilities under teacher forcing. Eq.~(\ref{eq:dpo_delta}) defines $\Delta(x_i,y_w,y_l)$ as the log-probability gap between the chosen and rejected responses after both are normalized by the reference model. A larger value of $\Delta$ means that the trainable  model prefers the chosen response more strongly than the rejected one  relative to the reference model. We initialize DPO from the RAFT model and train on the preference pairs. Together, RAFT provides structural supervision from retrieved Solidity examples, while DPO provides contrastive semantic supervision from SAP-based preference pairs.

\section{Experimental Setup}
\label{sec:evaluation_setup}

\subsection{Dataset}
\label{subsec:dataset}

To evaluate the effectiveness of our framework, we use \textbf{SolidityBench}~\cite{Chen2026RepositoryLevelSolidity}, a repository-level Solidity generation dataset constructed from three real-world Solidity sources: Synthetix, OpenZeppelin, and verified contracts on Etherscan. The dataset contains 5,470 Solidity contracts, each paired with a natural-language requirement. The source distribution is 550 samples from Synthetix, 1,150 from OpenZeppelin, and 3,770 from Etherscan. These sources provide complementary coverage: Synthetix contributes complex DeFi protocol code; OpenZeppelin contributes audited reusable contract components; and Etherscan contributes deployed contracts with diverse application logic and coding styles.

We split SolidityBench into training, validation, and test sets with an 8:1:1 ratio using random sampling with a fixed seed, yielding 4,376, 547, and 547 samples, respectively. The average contract contains 1,737.55 tokens and 19.09 functions, while the average requirement contains 61.68 tokens. These statistics reflect the complete-contract nature of the task: the model is required to generate not only local function bodies, but also declarations, imports, inheritance relations, modifiers, events, and shared state that coordinate multiple functions.

\subsection{Research Questions}
\label{subsec:research_questions}

Our experiments aim to answer the following four research questions.

\begin{itemize}

    \item \textbf{RQ1:} Does augmenting training with retrieval improve repository-level Solidity generation? This question examines whether providing the model with similar contracts retrieved via BM25 during fine-tuning provides benefits beyond standard SFT.
    \item \textbf{RQ2:} Which rejected-sample construction strategy is most effective for DPO? This question investigates whether Solidity-specific perturbations yield stronger preference signals than language-agnostic alternatives.
    \item \textbf{RQ3:} How much do RAFT and SAP-based DPO each contribute to the final pipeline? This question isolates the individual contributions of retrieval-augmented fine-tuning and preference optimization to the overall performance.
    \item \textbf{RQ4:} How sensitive is the framework to retrieval size $k$ and SAP perturbation ratio $\tau$? This question examines the impact of these parameters on generation quality and training stability.
\end{itemize}

\subsection{Evaluation Metrics}
\label{subsec:evaluation_metrics}

We evaluate the generated contracts using BLEU~\cite{Papineni2002BLEU} and SolidityScore (SS)~\cite{Chen2026RepositoryLevelSolidity}. The two metrics provide complementary offline signals for repository-level Solidity generation, where the model synthesizes a complete contract or contract library from a natural-language requirement.

BLEU measures surface-level overlap between a generated contract $\hat{y}$ and the reference contract $y^*$.
It captures lexical and structural similarity to the reference contract, including common declarations, function patterns, and statement-level fragments. However, Solidity contracts may remain textually similar while differing in access control, data handling, or value-transfer behavior. Therefore, BLEU alone is insufficient for assessing Solidity-specific generation quality.

To complement BLEU, we use SS, a Solidity-aware embedding-based similarity metric. Given a requirement $x$, a reference contract
$y^*=\langle y^*_1,\ldots,y^*_m\rangle$, and a generated contract
$\hat{y}=\langle \hat{y}_1,\ldots,\hat{y}_n\rangle$,
SS compares contextual token representations between $y^*$ and $\hat{y}$ under the same requirement context. Specifically, we encode the concatenated sequences $[x; y^*]$ and $[x; \hat{y}]$, and use only the contract-side token representations for matching.
Let $\mathbf{v}_{y^*_i}$ and $\mathbf{v}_{\hat{y}_j}$ denote the contextual embeddings of the $i$-th reference token and the $j$-th generated token, respectively. Token-level similarity is measured by cosine similarity.

SS  then computes domain-weighted recall and precision through greedy token matching. Let $M=|y^*|$ and $N=|\hat{y}|$ denote the token lengths of the reference and generated contracts, respectively.

\begin{equation}
R_{\mathrm{sol}}
:= \frac{ \sum_{i=1}^{M} w(y^*_i) \max_{j} \operatorname{sim}(\mathbf{v}_{y^*_i},\mathbf{v}_{\hat{y}_j}) }{ \sum_{i=1}^{M} w(y^*_i) },
\label{eq:ss_recall}
\end{equation}

\begin{equation}
P_{\mathrm{sol}}
:= \frac{ \sum_{j=1}^{N} w(\hat{y}_j) \max_{i} \operatorname{sim}(\mathbf{v}_{\hat{y}_j},\mathbf{v}_{y^*_i}) }{ \sum_{j=1}^{N} w(\hat{y}_j) }.
\label{eq:ss_precision}
\end{equation}

Here, $w(\cdot)$ is a domain-adaptive token weight derived from inverse document frequency (IDF) statistics computed solely on the Solidity training split~\cite{Robertson2004IDF}. It assigns higher importance to Solidity-informative tokens such as contract declarations, visibility modifiers,  data-location keywords, and value-transfer operations. $R_{\mathrm{sol}}$ measures how strongly reference-side tokens are covered by the generated code, while $P_{\mathrm{sol}}$ measures how closely generated tokens align with the reference. The final SS is the harmonic mean of these two terms:

\begin{equation}
SS = 2 \cdot \frac{P_{\mathrm{sol}} \cdot R_{\mathrm{sol}}}
{P_{\mathrm{sol}} + R_{\mathrm{sol}}}.
\label{eq:ss_f1}
\end{equation}

A higher SS indicates stronger Solidity-aware contextual alignment with the reference contract, especially on tokens and constructs that are important for smart contract behavior.

\subsection{Compared Methods}
\label{subsec:compared_methods}

We empirically evaluated four training configurations: \textbf{SFT} \cite{Chen2026RepositoryLevelSolidity}, \textbf{RAFT}, \textbf{SFT+DPO}, and \textbf{RAFT+DPO}. \textbf{SFT} fine-tunes the base model on requirement-contract pairs. \textbf{RAFT} uses the same target contracts as SFT for training but augments the input context with retrieved examples.
\textbf{SFT+DPO} applies preference optimization to a model initially trained with standard SFT using SAP-based rejected responses, thereby isolating the effect of preference data without retrieval-augmented initialization. \textbf{RAFT+DPO}, referred to as \textbf{RAPO-Sol}, is the full pipeline, where SAP-based DPO starts from the RAFT checkpoint. To ensure a fair comparison, all configurations use the same base models, data split, preprocessing, decoding procedure, and evaluation scripts.

\subsection{Evaluated Models}
 The evaluated backbones were CodeLlama-7B-Instruct-HF~\cite{Roziere2023CodeLlama}, DeepSeek-Coder-6.7B-Instruct~\cite{Guo2024DeepSeekCoder}, and Qwen2.5-Coder-7B-Instruct~\cite{Hui2024Qwen25Coder}, all obtained from their public Hugging Face releases. We selected these models for three reasons. First, they represent three distinct code-oriented LLM families with different pretraining data and training recipes, allowing us to examine whether the effectiveness of RAPO-Sol generalizes across backbones rather than depending on a single model family. Second, their comparable parameter scales (6.7B-7B) enable a controlled comparison under a similar training budget. Third, all three models provide publicly available weights, enabling the LoRA-based fine-tuning and DPO stages in our framework. In addition, these backbones were evaluated on the same repository-level Solidity generation setting in SolidityBench~\cite{Chen2026RepositoryLevelSolidity}, providing empirical grounding for their selection.

\subsection{Implementation Details}
\label{subsec:implementation_details}

All experiments were implemented in Python 3.12 with PyTorch 2.8.0, Hugging Face Transformers~\cite{Wolf2020Transformers}, and TRL~\cite{VonWerra2020TRL}. We used the \texttt{SFTTrainer} from TRL for SFT and RAFT, and the \texttt{DPOTrainer} for preference optimization.

Training was conducted on a single NVIDIA RTX~5090 GPU with 32\,GB VRAM. Since the backbone models are around the 7B scale and the inputs contain repository-level Solidity code, we used mixed precision and LoRA~\cite{Hu2022LoRA} for parameter-efficient fine-tuning. The original backbone parameters were frozen, and LoRA adapters were applied to the attention and feed-forward modules. We set the LoRA rank to~16, alpha to~32, and dropout to~0.05. The per-device batch size was~2 with eight gradient-accumulation steps. We used AdamW~\cite{Loshchilov2019AdamW}, a cosine learning-rate schedule, and a warm-up ratio of~0.05. Each SFT and RAFT model was trained for three epochs, and DPO was run for one epoch from the corresponding SFT or RAFT checkpoint. We did not set a separate \texttt{max\_steps}; the number of optimization steps was determined by the training-set size, batch size, and gradient accumulation.

For RAFT, retrieval was performed offline before fine-tuning. We built a BM25 index over the requirements in the training split and used each training requirement as the query. Retrieved examples were restricted to the training split, and the current sample was excluded from its own candidate list. To reduce retrieval leakage, we further filtered same-repository examples when repository metadata was available, removed contract-family variants, and discarded near duplicates after normalizing comments, whitespace, and identifiers. The main experiments used $k=1$. When the augmented input exceeded the maximum sequence length, we preserved the current requirement and target contract, and truncated retrieved examples from the end of the augmented context.

For DPO, the reference contract was used as the chosen response, and the rejected response was produced by one of the perturbation strategies described in Section~\ref{subsec:preference_data}. In the main setting, each reference contract was paired with one rejected response, so surface-level perturbation, general logical perturbation, and SAP produced preference datasets of the same size. For SAP, we fixed $|S|=1$ and set the perturbation ratio to $\tau=0.1$. Perturbed samples were discarded if they were unchanged, empty, removed the contract or library declaration, removed the target body, or failed the structural validity filter. The learning rate was $2\times10^{-5}$ for SFT and RAFT, and $5\times10^{-6}$ for DPO. The DPO preference strength was set to $\beta=0.1$.

During evaluation, all models were decoded from the natural-language requirement only. No retrieved examples were appended at test time, so the reported differences reflect changes introduced during training rather than additional inference context. We used greedy decoding to reduce generation variance. The main configuration used $k=1$ and $\tau=0.1$.

\section{Results and Analysis}
\label{sec:results}

\subsection{RQ1: Effect of Retrieval During Training}
\label{subsec:rq1}

To answer RQ1, we compare SFT and RAFT using the default retrieval size $k=1$. The results are shown in Table~\ref{table2_sft_vs_raft}. RAFT yields consistent improvements over SFT across all three backbone models and both evaluation metrics. For CodeLlama, RAFT improves BLEU by 3.669 points, from 28.173 to 31.842, and improves SS by 0.018, from 0.641 to 0.659. For DeepSeek-Coder, RAFT improves BLEU by 3.840 points, from 24.314 to 28.154, and improves SS by 0.029, from 0.609 to 0.638. Qwen2.5-Coder has the strongest SFT baseline, yet RAFT still improves BLEU by 2.762 points, from 35.962 to 38.724, and SS by 0.019, from 0.653 to 0.672. Overall, the positive changes across all six model–metric comparisons indicate that retrieval during training provides a consistent benefit for repository-level Solidity generation.

Because retrieved examples are not appended during inference, the gains cannot be attributed to extra prompt context. The results instead suggest that retrieved requirement-code examples provide useful structural cues during fine-tuning. These examples expose recurring Solidity patterns, such as access-control checks, event-emission logic, guarded state updates, and external-call handling. The improvement under requirement-only inference suggests that RAFT helps the model learn part of this structural signal during training, rather than relying on retrieval during inference.

\begin{table}[!htbp]
\centering
\caption{Performance comparison between SFT and RAFT with the default retrieval size $k=1$}
\label{table2_sft_vs_raft}
\footnotesize
\setlength{\tabcolsep}{5pt}
\renewcommand{\arraystretch}{1.08}
\begin{tabular}{llcc}
\toprule
Model & Method & BLEU & SS \\
\midrule
\multirow{2}{*}{CodeLlama-7B-Instruct-HF}
& SFT  & 28.173 & 0.641 \\
& RAFT & \textbf{31.842} & \textbf{0.659} \\
\midrule
\multirow{2}{*}{DeepSeek-Coder-6.7B-Instruct}
& SFT  & 24.314 & 0.609 \\
& RAFT & \textbf{28.154} & \textbf{0.638} \\
\midrule
\multirow{2}{*}{Qwen2.5-Coder-7B-Instruct}
& SFT  & 35.962 & 0.653 \\
& RAFT & \textbf{38.724} & \textbf{0.672} \\
\bottomrule
\end{tabular}
\end{table}

The gains are larger in BLEU than in SS, which is consistent with retrieved examples helping the model reproduce common lexical and structural patterns. SS also improves, but by a smaller margin. This pattern suggests that while retrieval during training strengthens structural learning, it does not directly teach the model to distinguish semantically correct code from near-miss variants. Preference optimization may therefore provide a complementary signal beyond retrieval during training.

\subsection{RQ2: Effect of Preference Data Construction}
\label{subsec:rq2}

To determine which rejected-sample construction strategy is most effective, we empirically compare three approaches under the same RAFT baseline. Table~\ref{table3_dpo_strategies} reports the results. The effect of DPO depends strongly on how the rejected responses are constructed. Surface-level perturbation yields small and inconsistent changes. It slightly improves BLEU for CodeLlama and DeepSeek-Coder, but reduces Qwen2.5-Coder's BLEU from 38.724 to 37.747. SS follows the same mixed pattern: it improves DeepSeek-Coder slightly, but decreases for CodeLlama and Qwen2.5-Coder.

\begin{table}[!htbp]
\centering
\caption{Effect of rejected-response construction strategies for DPO. Arrows mark changes relative to RAFT, and best results are in bold}
\label{table3_dpo_strategies}
\setlength{\tabcolsep}{2pt}
\renewcommand{\arraystretch}{1.08}
\resizebox{\columnwidth}{!}{
\begin{tabular}{lcccccc}
\toprule
\multirow{2}{*}{Training Configuration}
& \multicolumn{2}{c}{CodeLlama}
& \multicolumn{2}{c}{DeepSeek-Coder}
& \multicolumn{2}{c}{Qwen2.5-Coder} \\
\cmidrule(lr){2-3}
\cmidrule(lr){4-5}
\cmidrule(lr){6-7}
& BLEU & SS
& BLEU & SS
& BLEU & SS \\
\midrule
RAFT (Stage 1 Baseline)
& 31.842 & 0.659
& 28.154 & 0.638
& 38.724 & 0.672 \\

DPO Strategy 1: SLP
& 31.916\,$\uparrow$ & 0.654\,$\downarrow$
& 29.204\,$\uparrow$ & 0.642\,$\uparrow$
& 37.747\,$\downarrow$ & 0.668\,$\downarrow$ \\

DPO Strategy 2: GLP
& 32.163\,$\uparrow$ & 0.652\,$\downarrow$
& 28.512\,$\uparrow$ & 0.634\,$\downarrow$
& 38.982\,$\uparrow$ & 0.663\,$\downarrow$ \\

DPO Strategy 3: SAP
& \textbf{36.423}\,$\uparrow$ & \textbf{0.693}\,$\uparrow$
& \textbf{31.626}\,$\uparrow$ & \textbf{0.677}\,$\uparrow$
& \textbf{41.263}\,$\uparrow$ & \textbf{0.707}\,$\uparrow$ \\
\bottomrule
\end{tabular}
}
\end{table}

General logical perturbation yields a clearer BLEU gain but does not improve SS. For all three models, BLEU increases over the RAFT baseline, whereas SS decreases. For example, Qwen2.5-Coder reaches 38.982 BLEU, but its SS falls from 0.672 to 0.663. This split suggests that language-general mutations may improve surface-form similarity, but they do not consistently improve the Solidity-aware embedding signal captured by SS.

SAP is the only strategy that improves both metrics across all models. CodeLlama reaches 36.423 BLEU and 0.693 SS, DeepSeek-Coder reaches 31.626 and 0.677, and Qwen2.5-Coder reaches 41.263 and 0.707. Compared with the RAFT baseline, SAP-based DPO improves BLEU by 4.581, 3.472, and 2.539 points, and improves SS by 0.034, 0.039, and 0.035 for the three models, respectively. This consistency is important because SS reflects embedding-level semantic alignment that BLEU's surface $n$-gram overlap alone may miss.

The advantage of SAP stems from its domain-specific design: by targeting Solidity anchors related to validation, visibility, data handling, and value transfer, it creates hard negatives that surface perturbations and language-general mutations cannot. Surface perturbations provide only weak contrast, while language-general mutations miss the critical Solidity-specific dimensions that determine contract behavior.

\subsection{RQ3: Effect of the Training Pipeline}
\label{subsec:rq3}

\begin{figure}[!htbp]
    \centering
    \includegraphics[width=\linewidth]{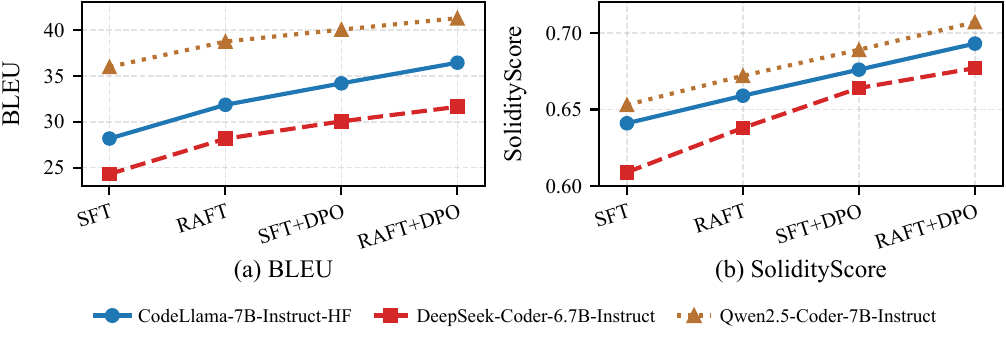}
    \caption{Ablation results of different training configurations across three backbone models.}
    \label{fig3_pipeline_ablation}
\end{figure}

To isolate the contribution of each stage, we compare four training configurations: SFT (the supervised baseline), RAFT, SFT+DPO, and RAFT+DPO (the full RAPO-Sol pipeline).  RAFT isolates the effect of training-time retrieval. SFT+DPO applies SAP-based DPO directly on the SFT checkpoint without retrieval-augmented initialization. RAFT+DPO combines both stages. The results are shown in Fig.~\ref{fig3_pipeline_ablation}.

From Fig.~\ref{fig3_pipeline_ablation}, it can be seen that both single-stage additions improve over SFT. RAFT improves all three models, as shown in RQ1. SFT+DPO also improves all three models, indicating that Solidity-aware preference data is beneficial even without retrieval-augmented initialization. On Qwen2.5-Coder, BLEU increases from 35.962 with SFT to 38.724 with RAFT and 40.024 with SFT+DPO. SS increases from 0.653 to 0.672 and 0.689. The same ordering appears for CodeLlama and DeepSeek-Coder.

The full RAFT+DPO pipeline achieves the highest score for every model and metric among the compared settings. CodeLlama reaches 36.423 BLEU and 0.693 SS, DeepSeek-Coder reaches 31.626 and 0.677, and Qwen2.5-Coder reaches 41.263 and 0.707. These scores are higher than both RAFT and SFT+DPO, suggesting that the two stages provide complementary benefits under our evaluation setting.

The ablation clarifies the role of each stage: RAFT improves exposure to contract-level structures, while SAP-based DPO encourages the model to assign higher preference to the chosen contract than to SAP-perturbed variants. The two stages therefore appear to address different weaknesses of SFT: RAFT provides the missing structural patterns, while DPO adds a contrastive signal that distinguishes correct code from Solidity-specific near misses.

\subsection{RQ4: Hyperparameter Sensitivity}
\label{subsec:rq4}

\subsubsection{Effect of Retrieval Size}
\label{subsubsec:retrieval_size}

To assess the sensitivity to the number of retrieved examples, we vary $k$ from 1 to 3 in RAFT. Table~\ref{table4_retrieval_size} reports the results. Moving from $k=1$ to $k=2$ yields only a marginal change, on the order of the second or third decimal place: CodeLlama moves from 31.842/0.659 to 31.993/0.664, DeepSeek-Coder from 28.154/0.638 to 28.378/0.642, and Qwen2.5-Coder from 38.724/0.672 to 38.852/0.674. The observed difference is small relative to the additional retrieved context.

\begin{table}[!htbp]
\centering
\caption{Sensitivity to retrieval size ($k$) in RAFT. Higher values are better; best results are in bold}
\label{table4_retrieval_size}
\setlength{\tabcolsep}{2pt}
\renewcommand{\arraystretch}{1.08}
\resizebox{\columnwidth}{!}{
\begin{tabular}{lcccccc}
\toprule
\multirow{2}{*}{Retrieval Size ($k$)}
& \multicolumn{2}{c}{CodeLlama-7B-Instruct-HF}
& \multicolumn{2}{c}{DeepSeek-Coder-6.7B-Instruct}
& \multicolumn{2}{c}{Qwen2.5-Coder-7B-Instruct} \\
\cmidrule(lr){2-3}
\cmidrule(lr){4-5}
\cmidrule(lr){6-7}
& BLEU $\uparrow$ & SS $\uparrow$
& BLEU $\uparrow$ & SS $\uparrow$
& BLEU $\uparrow$ & SS $\uparrow$ \\
\midrule
$k=1$ & 31.842 & 0.659 & 28.154 & 0.638 & 38.724 & 0.672 \\
$k=2$ & \textbf{31.993} & \textbf{0.664} & \textbf{28.378} & \textbf{0.642} & \textbf{38.852} & \textbf{0.674} \\
$k=3$ & 31.517 & 0.657 & 28.106 & 0.636 & 38.596 & 0.668 \\
\bottomrule
\end{tabular}
}
\end{table}

When $k=3$, performance drops for all models. The drop is modest but consistent: each backbone loses both BLEU and SS relative to $k=2$. More retrieved examples are therefore not necessarily better. Additional examples increase input length and may introduce redundant or weakly related patterns. For repository-level code, this cost is more visible because the requirement, retrieved contracts, and target code already compete for a limited context budget.

Because $k=2$ improves on $k=1$ only at the margin while doubling the retrieved context and the training cost, we adopt $k=1$ as the main setting. The framework is thus not tied to a single retrieval-size choice, and the trend across $k$ suggests that the quality and compactness of retrieved context matter more than its quantity.

\subsubsection{Effect of Perturbation Ratio}
\label{subsubsec:perturbation_ratio}

To assess the sensitivity to the perturbation ratio, we vary $\tau$ from 0.1 to 0.7. Table~\ref{table5_perturbation_ratio} reports the results. The best results are obtained at $\tau=0.1$ for all three models. As $\tau$ increases, both BLEU and SS decrease. For Qwen2.5-Coder, performance drops from 41.263/0.707 at $\tau=0.1$ to 40.694/0.704 at $\tau=0.3$, 39.123/0.683 at $\tau=0.5$, and 37.556/0.652 at $\tau=0.7$. CodeLlama follows the same pattern, falling from 36.423/0.693 to 31.052/0.651, and DeepSeek-Coder decreases from 31.626/0.677 to 27.854/0.643.

The monotonic drop suggests that preference-data quality is sensitive to perturbation strength. A small ratio changes only a few semantic anchors, so the chosen and rejected contracts remain close. The rejected code still resembles a plausible model output, but differs along Solidity-relevant anchors. A large ratio changes too many anchors at once, making the rejected code noisier and less realistic. This can weaken the preference signal.

Both retrieval size and perturbation strength thus call for moderation. These findings support the main configuration used in the paper, namely $k=1$ for efficient retrieval during training and $\tau=0.1$ for SAP-based preference data.
\begin{table}[!htbp]
\centering
\caption{Sensitivity to the SAP perturbation ratio ($\tau$). Higher values are better; best results are in bold}
\label{table5_perturbation_ratio}
\setlength{\tabcolsep}{2pt}
\renewcommand{\arraystretch}{1.08}
\resizebox{\columnwidth}{!}{
\begin{tabular}{lcccccc}
\toprule
\multirow{2}{*}{Ratio ($\tau$)}
& \multicolumn{2}{c}{CodeLlama-7B-Instruct-HF}
& \multicolumn{2}{c}{DeepSeek-Coder-6.7B-Instruct}
& \multicolumn{2}{c}{Qwen2.5-Coder-7B-Instruct} \\
\cmidrule(lr){2-3}
\cmidrule(lr){4-5}
\cmidrule(lr){6-7}
& BLEU $\uparrow$ & SS $\uparrow$
& BLEU $\uparrow$ & SS $\uparrow$
& BLEU $\uparrow$ & SS $\uparrow$ \\
\midrule
$\tau=0.1$ & \textbf{36.423} & \textbf{0.693} & \textbf{31.626} & \textbf{0.677} & \textbf{41.263} & \textbf{0.707} \\
$\tau=0.3$ & 35.958 & 0.688 & 31.104 & 0.672 & 40.694 & 0.704 \\
$\tau=0.5$ & 33.207 & 0.669 & 29.346 & 0.659 & 39.123 & 0.683 \\
$\tau=0.7$ & 31.052 & 0.651 & 27.854 & 0.643 & 37.556 & 0.652 \\
\bottomrule
\end{tabular}
}
\end{table}

\FloatBarrier

\section{Threats to Validity}
\label{sec:threats}

\textbf{Internal validity.}
All compared methods use the same base models, data split, preprocessing pipeline, decoding strategy, and evaluation scripts. Retrieved examples are fixed before training, and the current sample is excluded from retrieval. To minimize the risk of data leakage, we filter same-repository examples when metadata is available, discard near-duplicate code after normalizing comments, whitespace, and identifiers, and remove contracts sharing identical normalized structure but differing only in naming. These controls reduce variability from implementation choices, but they cannot fully eliminate the possibility that public code models may have observed some benchmark contracts during pretraining.

\textbf{Construct validity.}
BLEU measures surface-level $n$-gram overlap, while SolidityScore measures embedding-level semantic alignment. While higher scores generally correlate with higher-quality code, neither metric guarantees that a generated contract compiles, passes tests, or is secure~\cite{Papineni2002BLEU,Chen2021Evaluating,Ren2020CodeBLEU}. Execution-based metrics such as $\text{Pass}@k$ require a runnable host project, which is absent in our complete-contract generation setting. We therefore treat these offline metrics as comparative tools for controlled experimentation rather than as definitive measures of functional correctness. Future work could complement them with lightweight compiler diagnostics or targeted security checks.

\textbf{External validity.}
SolidityBench contains repository-level contracts drawn from three distinct sources, but it may not cover all contract types,  compiler-version conventions, or multi-contract protocols used in practice. The rejected responses used for DPO are generated by synthetic perturbation rules. While SAP targets security-critical Solidity anchors defined in Section~\ref{subsubsec:sap}, it may miss errors related to complex business logic, protocol-level invariants, or project-specific deployment assumptions that fall outside the scope of these anchors. Our results are based on 7B-level open-source code models; performance may change with larger models, alternative retrievers, or benchmarks that reconstruct complete repositories to enable execution-based evaluation~\cite{Peng2025SolEval,SolContractEval2025,Goel2026SmartEval}.
\section{Related Work}
\label{sec:related_work}

\subsection{Repository-Level Code Generation}
\label{subsec:repo_level_code_generation}

The task of repository-level code generation requires models to produce code that depends on context beyond a single function. Prior work has shown that this context can include imported files, class or contract definitions, project dependencies, and other cross-file information. For example, Shrivastava et al.~\cite{Shrivastava2023Repository} focus on selecting relevant context for code LLMs, CodePlan~\cite{Bairi2024CodePlan} studies planning for repository-level tasks, RepoCoder~\cite{Zhang2023RepoCoder} uses iterative retrieval and generation for code completion, and CrossCodeEval~\cite{Ding2023CrossCodeEval} provides a benchmark for cross-file completion.

These studies establish that project-level context matters, but most focus on general-purpose code or inference-time context construction. Solidity introduces additional constraints because contracts encode assets, permissions, gas-sensitive operations, and security-critical interactions. Recent benchmarks make these constraints explicit: SolEval~\cite{Peng2025SolEval} evaluates repository-level Solidity generation at the function level, while SolContractEval~\cite{SolContractEval2025} and SmartEval~\cite{Goel2026SmartEval} study contract-level or specification-driven generation. Our work is aligned with this line of evaluation but differs in both granularity and goal: we generate a complete contract or library directly from the requirement, and we focus on training rather than on inference-time context construction.

Compared with the above studies, the work closest to ours is that of Chen et al.~\cite{Chen2026RepositoryLevelSolidity}. Both works study repository-level Solidity code generation. While they conduct a comprehensive empirical study of various adaptation methods including zero-shot prompting, Chain-of-Thought reasoning, in-context learning, RAG, and SFT, and identify SFT as the most effective strategy for improving semantic correctness, we go beyond standard SFT by combining training-time retrieval with preference optimization to improve Solidity code generation.

\subsection{Retrieval-Augmented Training}
\label{subsec:retrieval_augmented_training}

RAG uses external evidence to improve model outputs. In natural language processing, RAG retrieves relevant documents and conditions generation on them at inference time~\cite{Lewis2020Retrieval}. In code generation, retrieval can expose examples, APIs, or project context that are absent from the immediate prompt. Repository prompt generation~\cite{Shrivastava2023Repository} and RepoCoder~\cite{Zhang2023RepoCoder} study ways to provide such context, while CrossCodeEval~\cite{Ding2023CrossCodeEval} evaluates models under cross-file code completion settings where this context is necessary.

Inference-time retrieval, however, changes the deployment setting. It requires a retrieval index, increases prompt length, and makes generation sensitive to retriever quality. RAFT moves retrieval into fine-tuning by adding retrieved examples during training rather than requiring them at test time~\cite{Zhang2024RAFT}. Recent work further combines retrieval-augmented fine-tuning with preference optimization for visual program generation~\cite{Kang2025Retrieval}. Our work follows this high-level direction but studies a different domain and failure mode. For repository-level Solidity, retrieved examples are used to teach recurring contract structures while the deployed model remains retrieval-free.

\subsection{Preference Optimization and Program Perturbation}
\label{subsec:preference_optimization}

Preference optimization trains models using paired comparisons between preferred and dispreferred outputs. Instruction tuning with human feedback shows that preference signals can align models with desired behavior~\cite{Ouyang2022Training}. DPO simplifies this process by directly optimizing on chosen and rejected responses without training a separate reward model~\cite{Rafailov2023Direct}. For code generation, the quality of the rejected response is especially important. If the rejected code differs only in style, the preference signal may be weak; if it is too heavily corrupted, the comparison may become trivial.

Prior work on program transformation and adversarial examples motivates controlled changes to code while keeping the program close to the original form~\cite{Yu2022Data,Yefet2020Adversarial}. Mutation testing provides a long-standing framework for creating behavior-changing variants of programs~\cite{Jia2011Analysis}, and recent work uses mutation-based tests to evaluate code LLMs~\cite{Li2024Mutation}. These works support the use of perturbed code as rejected responses, but they are not specific to Solidity.

Our main difference lies in the perturbation target. SAP builds rejected responses around the Solidity semantic anchors. These anchors are not arbitrary tokens; they are connected to validation, access control, data handling, and external interaction behavior. Their importance is well-documented: smart contract attack studies and vulnerability taxonomies identify these constructs as primary sources of critical flaws~\cite{Atzei2017Survey,Vidal2024OpenSCV,Ferreira2020SmartBugs}; static analyzers and fuzzers target them for vulnerability detection~\cite{Tsankov2018Securify,Tikhomirov2018SmartCheck,Jiang2018ContractFuzzer,Feist2019Slither,Brent2018Vandal}; and empirical tool evaluations confirm that errors around these constructs dominate real-world contract defects~\cite{Durieux2020Empirical}. We therefore leverage this domain knowledge to design SAP, which constructs Solidity-specific hard negatives that are close in surface form but semantically flawed along security-critical dimensions, in contrast to prior work that relies on language-agnostic or surface-level perturbations.

\section{Conclusion}
\label{sec:conclusion}

This paper presented RAPO-Sol, a two-stage framework for repository-level Solidity code generation that combines RAFT  with SAP-based DPO. RAFT helps the model learn recurring repository- and contract-level structures through retrieval during training while keeping inference retrieval-free. SAP-based DPO further improves semantic alignment by contrasting reference implementations with syntactically plausible but semantically perturbed hard negatives. Experiments on SolidityBench with three open-source code models show consistent improvements over standard SFT, with SAP-based DPO yielding additional gains and the full RAPO-Sol pipeline achieving the best BLEU and SolidityScore among the compared settings. These results suggest that training-time retrieval and Solidity-aware preference data provide complementary supervision for repository-level smart contract generation. Future work will explore stronger execution-aware feedback, such as incorporating compiler diagnostic messages as additional preference signals during DPO training. We also plan to investigate repository reconstruction and finer-grained semantic checks to better bridge the gap between offline metrics and functional correctness.

\bibliographystyle{IEEEtran}
\balance
\bibliography{citations.bib}

\end{document}